\documentclass{article}
\usepackage{amssymb}
\usepackage{pgfplots,tikz,comment}
\pgfplotsset{width=7cm,compat=1.18}
\usepackage{pgfplotstable}
\usepackage{float}
\usetikzlibrary{decorations.markings}
\tikzset
 {every pin/.style = {pin edge = {<-}}, 
  > = stealth, 
  flow/.style = 
   {decoration = {markings, mark=at position #1 with {\arrow{>}}},
    postaction = {decorate}
   },
  flow/.default = 0.5,   
  main/.style = {line width=1pt}
 }
\usepackage{amsmath}
\usepackage{amsfonts}
\usepackage{bm}
\usepackage{amsthm}
\usepackage{graphicx}
\usepackage{url,comment,xcolor}
\usepackage[normalem]{ulem}
\makeatletter
\def\squiggly{\bgroup \markoverwith{\textcolor{red}{\lower3.5\p@\hbox{\sixly \char58}}}\ULon}
\makeatother
\usepackage{bbm,mathrsfs}
\usepackage[unicode=true,psdextra,bookmarks=true,bookmarksnumbered=true,bookmarksopen=false,
 breaklinks=true,pdfborder={0 0 1},backref=false,colorlinks=true]
 {hyperref}
\usepackage{authblk}

\begin{document}

\title{A Kuramoto Model for the Bound State Aharonov-Bohm Effect}
\author[1]{Alviu Rey Nasir\footnote{Corresponding author.  E-mail: alviurey.nasir@g.msuiit.edu.ph}}

\author[2]{Jos\'{e}~Lu\'{i}s~Da~Silva}
\author[1]{Jingle~Magallanes}
\author[3]{Herry~Pribawanto~Suryawan}
\author[1, 4]{Roshin Marielle Nasir-Britos}

\affil[1]{Department of Physics, College of Science and Mathematics\newline \& Premier Research Institute of Science and Mathematics\newline Mindanao State University-Iligan Institute of Technology\newline Iligan City 9200, Philippines}

\affil[2]{CIMA Faculdade de C{\^e}ncias Exatas e da Engenharia\newline Campus Universit{\'a}rio da Penteada, Universidade da Madeira\newline 9020-105 Funchal, Madeira, Portugal}

\affil[3]{Department of Mathematics, Faculty of Science and Technology\newline Sanata Dharma University, Yogyakarta 55283, Indonesia}

\affil[4]{Department of General Education, Initao College\newline Initao, Misamis Oriental, 9022 Philippines}
\maketitle

\begin{abstract}
Starting with the overall wave function expression for the electron in an Aharonov-Bohm potential, we derive a version of the Kuramoto Model describing the phase dynamics of the bound state of the quantum mechanical system.
\end{abstract}

\section{Introduction and Motivation}

The Aharonov-Bohm (AB) effect \cite{aharonov1959significance}, which is an interference phenomenon, continues to exhibit quantum mechanical features that have not yet been fully understood; see, for example, Refs.~\cite{batelaan2009aharonov, cembranos2024aharonov, biswas2024anomalous, wakamatsu2024revisiting}.  On the other hand, the Kuramoto model (KM) \cite{kuramoto1975self} has been applied to many branches of science (see review papers, e.g., Ref.~\cite{acebron2005kuramoto, rodrigues2016kuramoto, guo2021overviews}) but scarcely to quantum mechanics. For quantum applications of the KM, see, examples given in Refs.~\cite{delmonte2023quantum, bastidas2024quantum, de2013quantum, morrison2013quantum}.
Describing the AB effect from the KM's perspective is therefore of interest not considered elsewhere.

Synchronizations in quantum systems have recently been studied based on either having a classical analog (see, e.g., Refs.~\cite{lee2013quantum, walter2015quantum}) or none at all; see, e.g., Ref.~\cite{roulet2018quantum}.  Furthermore, synchronizations of interfering particles, like between photons (see, e.g., Ref.~\cite{makino2016synchronization}) and between electrons (see, e.g., Ref.~\cite{neder2007interference}) have also been considered.  Since the synchronization condition is a main characteristic of the KM, it is therefore of particular interest to use the model to study the synchronization conditions of the bound state AB effect and their relationship with the interference occurring within the quantum system.  Alongside the synchronization conditions, we are also interested in determining the extent of the magnetic field region near the interference line of the AB effect.

In this letter, beginning with the general case of the AB effect \cite{aharonov1959significance}, we consider its simplest case, the bound state, by constraining the path taken by the electron to that of a circle with constant radius; see, e.g., Refs.~\cite{peshkin1990aharonov, bernido2002path, kretzschmar1965aharonov}.  Then, with its known interference phase difference and wave functions (see, e.g., Ref.\cite{aharonov1959significance}), we propose a version of the KM and study the phase dynamics of the system. 

\section{A Kuramoto Model for the System}

Begin with a KM model of the form \cite{kuramoto1975self}
\begin{equation}
    \dot{\Theta}_i := \frac{\mathrm{d}\Theta_i}{\mathrm{d}t} = \omega_i + \frac{K}{N} \sum_{j=1}^N \sin \left( \Theta_j - \Theta_i \right),
    \label{km}
\end{equation}
where $\Theta_i = \Theta_i(t)$ is the phase of the phase oscillator $i$ (with $i=1, 2, \ldots, N$) as a function of time $t \geq 0$,  $\omega_i$ its angular frequency, $K$ a uniform coupling strength coefficient, and $N$ the total number of phase oscillators.

\begin{figure}[ht] 
\caption{\label{fig:solenoid1}A top-view illustration of the paths taken by a spinning electron around the solenoid from which came the magnetic flux ``inducing'' Aharonov-Bohm effect.}
\centering
 \centering 
\usetikzlibrary {shapes.geometric,arrows.meta} 
\begin{tikzpicture}[thick,>=stealth]  
\draw[arrows={->[scale=.9]},red](3.7,0)--(3.2,0);
\draw[arrows={->[slant=-.1,scale=.8]},red](1.7,.2) arc [start angle=0, end angle=36, radius=20pt];
\draw[arrows={->[slant=-.1,scale=.8]},red](1.7,-.2) arc [start angle=0, end angle=-36, radius=20pt];
\draw (2.2,.8) node {\footnotesize \textit{ccw (Path 1)}}; 
\draw (2.2,-.8) node {\footnotesize \textit{cw (Path 2)}}; 
\draw[->](4.5,-.5)--(4.1,-.1);
\draw (4.5,-.7) node {\scriptsize Charged};
\draw (4.5,-.9) node {\scriptsize Particle};
\draw (4.5,-1.2) node {\scriptsize (two of such)};
\filldraw [very thick,draw=black!50,fill=black!20,opacity=7.7] (4,0) circle [radius=4pt]; 
\draw (.25,-0.30) node {$\phi$}; 
\draw (.2,0.65) node {$\theta$}; 
\filldraw [very thick,draw=blue!50,fill=blue!20,opacity=0.8] (0,0) circle [radius=7pt];
\draw[arrows={->[slant=-0.3,scale=.8]},black](0.5,0) arc [start angle=0, end angle=135, radius=14pt];
\draw[purple,style=dashed](0,0)--(-1.18,0.95) node [above, pos=.50, above,color=black] {$R$};
\draw[purple,style=dashed](0,0)--(-0.18,-0.18) node [below, pos=.8, below,color=black] {$R_0$};
\draw[blue,style=dashed](1.5,0)--(-1.5,0);

\draw[red,style=dashed](-2,0)--(-1.5,0);
\draw[red,style=dashed](1.5,0)--(3,0);
\begin{scope}
    \clip (-2,0) rectangle (2,2);
    \draw (0,0) circle(1.5);
\end{scope}
\begin{scope}
    \clip (-2,0) rectangle (2,-2);
    \draw[blue,style=dashed] (0,0) circle(1.5);
\end{scope}
\end{tikzpicture}
\end{figure}
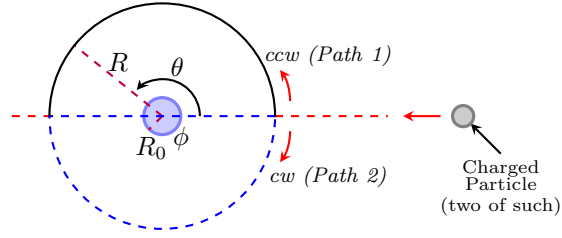

Consider now a two-electron system, with each electron initially at the angular position $\theta_0 := \theta(0) = 0$ moving along a circular path of constant radius $R,$ one in the upper semicircle and the other in the lower semicircle; see Fig.~\ref{fig:solenoid1}. The two paths, namely, Path 1 and Path 2, form a circular ring except at the endpoint $\theta(t) = \pi$.

In the center of the ring, there is an AB effect source, a magnetically impenetrable solenoid of radius $R_0$ oriented perpendicular to the semicircular paths and generating a magnetic field of flux $\phi$. The resulting magnetic field lines emerge from the top, flow down the sides of the solenoid beyond the trajectories, and return through the bottom. Since this magnetic field does not intersect the paths of the electrons, the paths (of radius $R \gg R_0$) of the electrons are in a region where the magnetic field is excluded. Consequently, there is a phase shift between the electron trajectories at their mutual endpoint, specifically at $\theta(t) = \pi$.

In the general case where the ``detecting" particles (i.e., the charged particles moving along the semicircular paths that will sense the AB effect) having charge $le, \, l \in \mathbb{Z} \setminus \{0\}$, the phase difference between Path 1 and Path 2, when they interfere at $\theta(t) = \pi$, is \cite{aharonov1959significance, peshkin1990aharonov}
\begin{equation}
    \Theta_1 - \Theta_2  = \frac{S_1 - S_2}{\hbar} =  2\pi\alpha, \qquad \alpha := \frac{le \phi}{2\pi \hbar c};
   \label{phase}
\end{equation}
where $S_i$ represents the action of the classical path $i$, $\hbar$ the reduced Planck's constant, and $c$ the speed of light in vacuum.  Note that the quantity $\alpha$, known as the flux parameter, only takes specific discrete values of $\phi = (\phi_0/ n), \, n \in \mathbb{Z} \setminus \{0\}$, where $\phi_0 := 2\pi \hbar c/ e$ (known as the quantum of flux or London's unit) and $ne$ is the charge used by the ``source" of the magnetic flux in the solenoid.  In particular, with $\alpha = l/n$, the AB effect is detectable only for $\{l/n\}$, where $\{ \, \cdot \, \}$ represents the fractional part of the number.  In our case, where $l=-1$, the AB effect is detectable whenever $|n| \neq 1$. One can find a detailed theoretical description in Refs.~\cite{aharonov1959significance, peshkin1990aharonov} and some supporting experiments in the works \cite{peng2010aharonov} and \cite{xiu2011manipulating}.

Hence, with $N=2$, i.e., for Paths~1 and 2 in Fig.~1, Eq.~(\ref{km}) gives
\begin{equation}
    \dot{\Theta}_1 = \omega_1 + \frac{K}{2} \sin \left( {\Theta}_2 - {\Theta}_1 \right),
    \label{eq:km-ab}
\end{equation}
and
\begin{equation}
    \dot{\Theta}_2 = \omega_2 - \frac{K}{2} \sin \left( {\Theta}_2 - {\Theta}_1 \right).
    \label{eq:km-ab2}
\end{equation}

This study aims to determine $\omega_1$, $\omega_2$, and $K$.  To do so, we make the following assumptions:
\begin{enumerate}
    \item the probability for an electron to take either of the two paths is the same;
    \item the two electrons simultaneously start at $\theta_1(t_0)=\theta_2(t_0)=0$, one then proceeds towards the end of the semicircular track through Path 1 and the other through Path 2;
    \item the electrons travel the same amount of angular displacement $\theta := \theta(t)$ (differing only in sign of $\theta$), at time interval $t - t_0 = t,$
    thus enabling them to interfere as $|\theta(t)| \rightarrow \pi$.
\end{enumerate}

Now, in quantum mechanics, the wave function $\psi(t)$ for the (bound state) Aharonov Bohm effect can be expressed in terms of $S_1$ and $S_2$ as \cite{aharonov1959significance}
\begin{equation}
    \psi(t) = \psi_1^0(t) \mathrm{e}^{-(\mathrm{i}/\hbar)S_1} + \psi_2^0(t) \mathrm{e}^{-(\mathrm{i}/\hbar)S_2}, \label{eq:psi(t)in_terms_of_S}
\end{equation}
where $\psi_i^0$ denotes the wave function for the free particle moving along path $i$. Furthermore, since this system also undergoes a scattering process \cite{aharonov1959significance}, then $\psi$ can further be expressed in terms of the incident wave function $\psi_\mathrm{inc}$ and the scattered wave function $\psi_{\mathrm{scatt}}$ as \cite{aharonov1959significance}
\begin{equation}
    \psi(t) = \psi_{\mathrm{inc}}(t) + \psi_{\mathrm{scatt}}(t), \label{eq:psi(t)in_terms_of_inc_and_scattered}
\end{equation}
where, for $ |\theta| < \pi \Leftrightarrow |\theta(t)| < \pi$, $\forall t\geq0$,
\begin{equation}
    \psi_{\mathrm{inc}}(t) := \mathrm{e}^{-\mathrm{i}\left( \alpha \theta(t) + R k \cos (\theta(t)) \right)}, \label{eq:psi_inc2}
\end{equation}
and
\begin{equation}
    \psi_{\mathrm{scatt}}(t) := \frac{\sin (\pi \alpha)}{\sqrt{2\pi \mathrm{i} Rk}\cos (\theta(t)/2)}\mathrm{e}^{-\mathrm{i}\left(\frac{\theta(t)}{2} - R k\right)}, \label{eq:psi_scattered}
\end{equation}
with $k \in \mathbb{R}^+$ being the magnitude of the wave vector of the incident electron.

Let
\begin{equation}
    \psi_{\mathrm{inc}}(t) = A(t)\mathrm{e}^{-\mathrm{i}\int_0^t \omega_\mathrm{inc}(s)\,\mathrm{d}s}, \label{eq:psi_inc_general}
\end{equation}
and
\begin{eqnarray}
    \psi_{\mathrm{scatt}}(t) = B(t)\mathrm{e}^{-\mathrm{i}\int_0^t \omega_\mathrm{scatt}(s)\,\mathrm{d}s}, \quad A(t), B(t) \in \mathbb{R}, \label{eq:psi_scatt_general}
\end{eqnarray}
where $\omega_\mathrm{inc} := \omega_\mathrm{inc}(t)$ and $\omega_\mathrm{scatt} := \omega_\mathrm{scatt}(t)$ are a form of phase frequencies (i.e., having the dimensions of phase divided by time) for the incident and the scattered waves, respectively.  Then, upon inspecting Eqs.~(\ref{phase})---(\ref{eq:psi_scatt_general}), we propose that $\omega_1$ and $\omega_2$ can be obtained as having the same expression but different in sign of $\theta$ due to Path 1 (ccw) and Path 2 (cw), namely, $\omega_1 = \omega_\mathrm{inc}$ for $0 \leq \theta < \pi$ (let $\omega_\mathrm{inc}^+ := \omega_1$), and $\omega_2 = \omega_\mathrm{inc}$ for $-\pi < \theta \leq 0$ (let $\omega_\mathrm{inc}^- := \omega_2$).  Therefore,
\begin{equation}
    \int_0^t \omega_\mathrm{inc}(s)\,\mathrm{d}s = \alpha \theta(t) + R k\cos (\theta(t)),\nonumber \label{eq:int_of_omega}
\end{equation}
from which follows
\begin{eqnarray}
    \omega_\mathrm{inc}(t) &=& \frac{\mathrm{d}}{\mathrm{d}t} \big( \alpha \theta(t) + R k \cos(\theta(t)) \big) \nonumber\\
    &=& \big(\alpha - R k \sin (\theta(t))\big) \dot{\theta}(t). \label{eq:omega}
\end{eqnarray}
Similarly, we are led to model $K$ in $\omega_\mathrm{scatt}$.  Hence,
\begin{equation}
    K = K(\theta(t),\dot{\theta}(t))
    = \frac{\mathrm{d}}{\mathrm{d}t} \left( \frac{\theta(t)}{2} -R k + \frac{\pi}{4}\right) = \frac{\dot{\theta}(t)}{2}. \label{eq:k/2}
\end{equation}
Notice that we have excluded the terms arising from the prefactor on the right-hand side of Eq.~(\ref{eq:psi_scattered}), which would contribute to an imaginary phase $z$ when represented as $e^{-\mathrm{i}z}$.  In other words, we have not included the real term $B(t)$ in modeling $K$.

With Eqs.~(\ref{km}), (\ref{eq:omega}) and (\ref{eq:k/2}), Eqs.~(\ref{eq:km-ab}) and (\ref{eq:km-ab2}) now read, respectively (for $0 \leq \theta < \pi$),
\begin{equation}
 \dot{\Theta}_1 =  \left[ \alpha  - R k\sin (\theta)  + \frac{1}{2}\sin \left( \Theta_2 - \Theta_1 \right) \right]\dot{\theta},\label{eq:thetadot1}
\end{equation}
and (for $-\pi < \theta \leq 0$)
\begin{equation}
 \dot{\Theta}_2 =  \left[ \alpha  - R k\sin (\theta)  - \frac{1}{2}\sin \left( \Theta_2 - \Theta_1 \right) \right]\dot{\theta}. \label{eq:thetadot2}
\end{equation}
Note that, in the case of Path 2 in Fig.~1, $\dot{\theta} < 0 $ is the angular velocity in the clockwise (cw) direction.

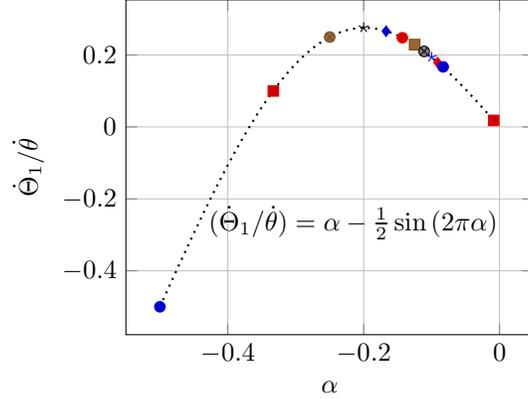
\begin{figure}[ht] 
\caption{\label{fig:plot1}The plot of $\dot{\Theta}_1/\dot{\theta}$ versus $\alpha$ as $\theta \rightarrow \pi$ for $\dot{\theta}\neq 0$ (see values in Table~\ref{tab:1}).}
\centering
\begin{tikzpicture}
    \begin{axis}[legend pos=outer north east, xlabel={$\alpha$}, ylabel={$\dot{\Theta}_1/\dot{\theta}$}, legend pos=outer north east, legend style={at={(axis cs:0.62,1.1)},anchor=north},legend cell align=left, grid=both, grid style={line width=.1pt, draw=gray!10}, major grid style={line width=.2pt,draw=gray!50}]
    \addplot coordinates {( -0.5, -0.5)}; 
    \addplot coordinates {( -0.333, 0.1)}; 
    \addplot coordinates {( -0.25, 0.25)}; 
    \addplot coordinates {( -0.2, 0.276)}; 
    \addplot coordinates {( -0.167, 0.266)}; 
    \addplot coordinates {( -0.143, 0.248)}; 
    \addplot coordinates {( -0.125, 0.229)}; 
    \addplot coordinates {( -0.111, 0.21)}; 
    \addplot coordinates {( -0.1, 0.194)};
    \addplot coordinates {( -0.091, 0.179)};
    \addplot coordinates {( -0.083, 0.167)};
    \addplot coordinates {( -0.00848, 0.018)};
    \addplot [pin distance=10mm,dotted, domain=-0.5:0,samples=250, thick, black ] {(x-0.5*(sin(2*(180/pi)*pi*x)))} node[pos=0.2, right]{$(\dot{\Theta}_1/\dot{\theta})=\alpha - \frac{1}{2}\sin \left( 2\pi \alpha \right)$};
    \end{axis}
\end{tikzpicture}
\end{figure}

At this point, using Eqs.~(\ref{eq:thetadot1}) and (\ref{eq:thetadot2}), we obtain for $0 \leq \theta < \pi$
\begin{equation}
    \Theta_2 - \Theta_1  = -2 \theta \alpha , \label{eq:alphapi}
\end{equation}
and for $-\pi < \theta \leq 0$
\begin{equation}
    \Theta_2 - \Theta_1  = 2 \theta \alpha, \label{eq:alphapi2}
\end{equation}
which confirms the results of Ref.~\cite{aharonov1959significance} (see Eq.~(\ref{phase})) as $|\theta|~\rightarrow~\pi$.  Therefore (for $0 \leq \theta < \pi$),
\begin{equation}
 \lim_{\theta \rightarrow \pi}\dot{\Theta}_1 = \left[ \alpha - \frac{1}{2}\sin \left( 2\pi \alpha \right) \right]\dot{\theta}, \label{eq:thetadot1untilpiatpi}
\end{equation}
and (for $-\pi < \theta \leq 0$)
\begin{equation}
 \lim_{\theta \rightarrow -\pi}\dot{\Theta}_2 = \left[ \alpha  + \frac{1}{2}\sin \left( 2\pi \alpha \right) \right]\dot{\theta}, \label{eq:thetadot2untilpiatpi}
\end{equation}
with which we list down on Table~\ref{tab:plot1} the corresponding values of $\dot{\Theta}_1/\dot{\theta}$ (for $\dot{\theta} \neq 0$) for different values of $\alpha$.  We plot these numbers on a diagram in Fig.~\ref{fig:plot1}.

\begin{table}
\caption{\label{tab:plot1} AB Effect with the electron as the detecting particle as $\theta \rightarrow \pi$ for $\dot{\theta} \neq 0$ with different flux parameter values, $\alpha$.}
\label{tab:1}
\begin{center}
\begin{tabular}{lllll}
$\alpha$ & $n$ & Similar Element& $\dot{\Theta}_1/\dot{\theta}$ & AB Effect?\\ \hline
$-1$ & 1 & H & -1.0 & No \\
$-\frac{1}{2}$ & 2 & He & -0.5 & Yes \\[.1cm]
$-\frac{1}{3}$ & 3 & Li & 0.1 & Yes \\[.1cm]
$-\frac{1}{4}$ & 4 & Be & 0.250 & Yes \\[.1cm]
$-\frac{1}{5}$ & 5 & B & 0.276 & Yes \\[.1cm]
$-\frac{1}{6}$ & 6 & C & 0.266 & Yes \\[.1cm]
$-\frac{1}{7}$ & 7 & N & 0.248 & Yes \\[.1cm]
$-\frac{1}{8}$ & 8 & O & 0.229 & Yes \\[.1cm]
$-\frac{1}{9}$ & 9 & F & 0.21 & Yes \\[.1cm]
$-\frac{1}{10}$ & 10 & Ne & 0.194 & Yes \\[.1cm]
$-\frac{1}{11}$ & 11 & Na & 0.179 & Yes \\[.1cm]
$-\frac{1}{12}$ & 12 & Mg & 0.167 & Yes \\
[.1cm]
$\vdots$ & $\vdots$ & $\vdots$ &  $\vdots$ & $\vdots$ \\
[.1cm]
$-\frac{1}{118}$ & 118 & Og & 0.018 & Yes
\end{tabular}
\end{center}
\end{table}

Note here that, at the interference point (in polar coordinates $(r,\theta) = (R,\pi)$), the phase of each of the electrons is independent of the circular path radius.  However, it is essential to keep in mind that the circle's radius must meet the constraint to ensure the paths remain within the magnetically excluded region. Therefore, we must further analyze our KM to determine the ``critical" radius $R_\mathrm{crit}$. This radius is likely related to the radius through which the magnetic field lines pass, i.e., no longer magnetically excluded.

\begin{figure}[ht] 
\caption{\label{fig:plot2}The plot of $\dot{\Theta}_1/\dot{\theta}$ versus $\theta$ for $\dot{\theta} \neq 0$ (see Eq.~(\ref{eq:thetadot1withthetaalpha})) for different values of $Rk$ when $\alpha = -1/2$.}
\centering
\begin{tikzpicture}[circ/.style={shape=circle, inner sep=1pt, draw, node contents=}]
    \begin{axis}[domain=0:179, legend style={cells={align=left},at={(axis cs:185,-1.35)},anchor=south west}, xlabel={$\theta$},
    ylabel={$\dot{\Theta}_1/\dot{\theta}$}, no marks, every axis plot post/.append style={very thick}, grid=both, grid style={line width=.1pt, draw=gray!10}, major grid style={line width=.2pt,draw=gray!50}]
    \addplot {(-0.5-0.0001*sin(x)+0.5*sin(2*x*0.5))};
    \addplot {(-0.5-0.02*sin(x)+0.5*sin(2*x*0.5))};
    \addplot {(-0.5-0.04*sin(x)+0.5*sin(2*x*0.5))};
    \addplot {(-0.5-0.1*sin(x)+0.5*sin(2*x*0.5))};
    \addplot {(-0.5-0.2*sin(x)+0.5*sin(2*x*0.5))};
    \addplot {(-0.5-0.3*sin(x)+0.5*sin(2*x*0.5))};
    \addplot {(-0.5-0.4*sin(x)+0.5*sin(2*x*0.5))};
    \addplot {(-0.5-0.45*sin(x)+0.5*sin(2*x*0.5))};
    \addplot {(-0.5-0.5*sin(x)+0.5*sin(2*x*0.5))};
    \addplot {(-0.5-0.55*sin(x)+0.5*sin(2*x*0.5))};
    \addplot {(-0.5-0.6*sin(x)+0.5*sin(2*x*0.5))};
    \addplot {(-0.5-0.7*sin(x)+0.5*sin(2*x*0.5))};
    \addplot {(-0.5-1*sin(x)+0.5*sin(2*x*0.5))};
    \legend{$Rk=$\\$0.0001$,$0.02$,$0.04$, $0.1$, $0.2$, $0.3$, $0.4$, $0.45$, $0.5$, $0.55$, $0.6$, $0.7$, $1.0$};
    \end{axis}
\end{tikzpicture}
\end{figure}
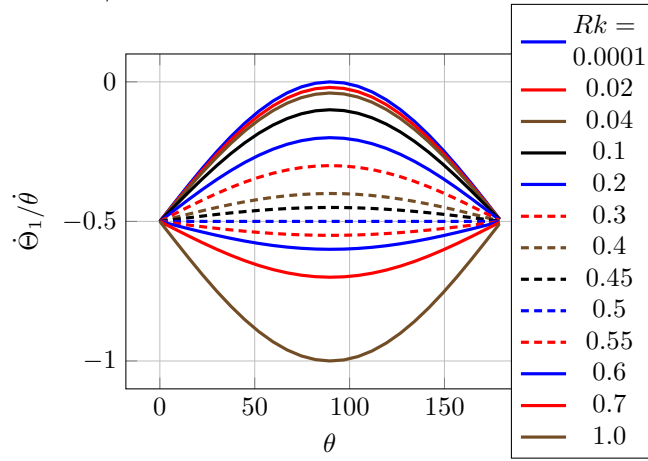

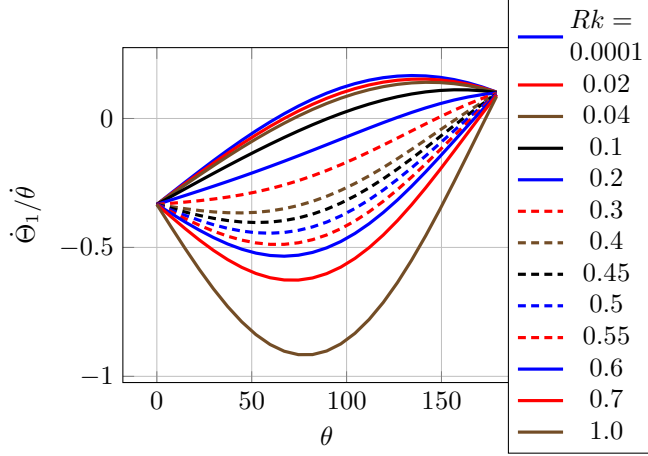
\begin{figure}[ht] 
\caption{\label{fig:plot2b}The plot of $\dot{\Theta}_1/\dot{\theta}$ versus $\theta$ for $\dot{\theta} \neq 0$ (see Eq.~(\ref{eq:thetadot1withthetaalpha})) for different values of $Rk$ when $\alpha = -1/3$.}
\centering
\begin{tikzpicture}[circ/.style={shape=circle, inner sep=1pt, draw, node contents=}]
    \begin{axis}[domain=0:179, legend style={cells={align=left},at={(axis cs:185,-1.3)},anchor=south west}, xlabel={$\theta$}, ylabel={$\dot{\Theta}_1/\dot{\theta}$}, no marks, every axis plot post/.append style={very thick}, grid=both, grid style={line width=.1pt, draw=gray!10}, major grid style={line width=.2pt,draw=gray!50}]
    \addplot {(-(1/3)-0.0001*sin(x)+0.5*sin(2*x*(1/3)))};
    \addplot {(-(1/3)-0.02*sin(x)+0.5*sin(2*x*(1/3)))};
    \addplot {(-(1/3)-0.04*sin(x)+0.5*sin(2*x*(1/3)))};
    \addplot {(-(1/3)-0.1*sin(x)+0.5*sin(2*x*(1/3)))};
    \addplot {(-(1/3)-0.2*sin(x)+0.5*sin(2*x*(1/3)))};
    \addplot {(-(1/3)-0.3*sin(x)+0.5*sin(2*x*(1/3)))};
    \addplot {(-(1/3)-0.4*sin(x)+0.5*sin(2*x*(1/3)))};
    \addplot {(-(1/3)-0.45*sin(x)+0.5*sin(2*x*(1/3)))};
    \addplot {(-(1/3)-0.5*sin(x)+0.5*sin(2*x*(1/3)))};
    \addplot {(-(1/3)-0.55*sin(x)+0.5*sin(2*x*(1/3)))};
    \addplot {(-(1/3)-0.6*sin(x)+0.5*sin(2*x*(1/3)))};
    \addplot {(-(1/3)-0.7*sin(x)+0.5*sin(2*x*(1/3)))};
    \addplot {(-(1/3)-1.0*sin(x)+0.5*sin(2*x*(1/3)))};
    \legend{$Rk=$\\$0.0001$,$0.02$,$0.04$, $0.1$, $0.2$, $0.3$, $0.4$, $0.45$, $0.5$, $0.55$, $0.6$, $0.7$, $1.0$}
    \end{axis}
\end{tikzpicture}
\end{figure}

Now, looking at Eqs.~(\ref{phase}), (\ref{eq:thetadot1}), (\ref{eq:thetadot2}), (\ref{eq:alphapi}) and (\ref{eq:alphapi2}), we find (for $0 \leq \theta < \pi$)
\begin{equation}
 \dot{\Theta}_1 = \left[ \alpha  - R k\sin (\theta)  + \frac{1}{2}\sin \left(-2\theta \alpha \right) \right]\dot{\theta},\label{eq:thetadot1withthetaalpha}
\end{equation}
and (for $-\pi < \theta \leq 0$)
\begin{equation}
 \dot{\Theta}_2 = \left[ \alpha  - R k\sin (\theta)  + \frac{1}{2}\sin \left(-2\theta \alpha \right) \right]\dot{\theta},\label{eq:thetadot2withthetaalpha}
\end{equation}
with which we plot the values of $\dot{\Theta}_1/\dot{\theta}$ versus $\theta$ for $\dot{\theta} \neq 0$ on a diagram in Fig.~\ref{fig:plot2} for $\alpha=-1/2,$ and in Fig.~\ref{fig:plot2b} for $\alpha=-1/3.$

We need to find out the range of values for $R$ within which we can detect the AB effect. This depends on the magnetic field and how much $R$ is within the excluded region. A more precise question would be: with a given flux parameter $\alpha$, how can we determine the maximum or critical $R$ to ensure that it remains in the exclusive region?

Looking at the graph in Fig.~\ref{fig:plot2}, one can see that the function $\dot{\Theta}_1/\dot{\theta}$ is a concave parabola downward (i.e., the extreme is a maximum) until a certain value $Rk = R_\mathrm{crit}k$, when it becomes a horizontal line.  After this, i.e., when $R > R_\mathrm{crit}$, increasing $Rk$ would make the function concave upward, that is, the extreme is a minimum.  On the other hand, a similar feature can be observed in the graph in Fig.~\ref{fig:plot2b} but slanting upward (from the left) in a pattern instead of just symmetric along the horizontal. 

We propose that for $\alpha = -1/2$, together with a given value of $k$, $R_\mathrm{crit}$ entails a unique radius $R$, at which the bound state AB effect cannot be detected, that is, the magnetic field lines pass through this radius, thus not magnetically excluded. Beyond $R_\mathrm{crit}$, the bound state AB effect resumes as there are no magnetic field lines in the region once again.  Although for $\alpha \neq -1/2$, for example, $\alpha = -1/3$, the function $\dot{\Theta}_1/\dot{\theta}$ is not as symmetrical as that for $\alpha = -1/2$, however, the critical radius, $R_\mathrm{crit}$, can still indicate the ``flipping" behavior, which is only attributable to the presence of the magnetic field lines.

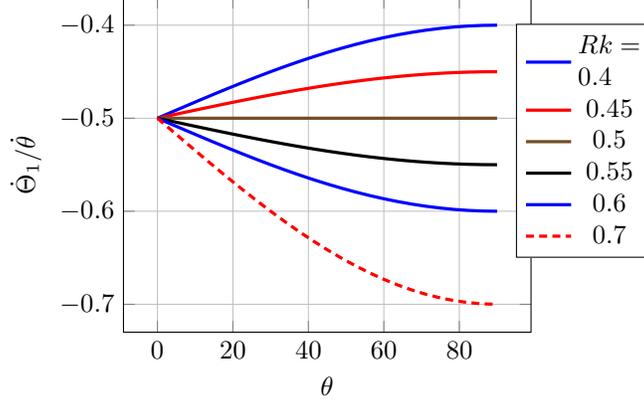
\begin{figure}[ht] 
\caption{\label{fig:3} Finding $R_\mathrm{crit}$ in the plot of $\dot{\Theta}_1/\dot{\theta}$ versus $\theta$ for $\dot{\theta} \neq 0$ for different values of $Rk$ when $\alpha = -1/2$.}
\centering
\begin{tikzpicture}
    \begin{axis}[domain=0:90, legend style={cells={align=left},at={(axis cs:95,-0.65)},anchor=south west}, xlabel={$\theta$}, ylabel={$\dot{\Theta}_1/\dot{\theta}$}, no marks, every axis plot post/.append style={very thick}, grid=both,
    grid style={line width=.1pt, draw=gray!10}, major grid style={line width=.2pt,draw=gray!50}, y tick label style={
        /pgf/number format/.cd, fixed, fixed zerofill, precision=1, /tikz/.cd}]
    \addplot {(-0.5-0.4*sin(x)+0.5*sin(2*x*0.5))};
    \addplot {(-0.5-0.45*sin(x)+0.5*sin(2*x*0.5))};
    \addplot {(-0.5-0.5*sin(x)+0.5*sin(2*x*0.5))};
    \addplot {(-0.5-0.55*sin(x)+0.5*sin(2*x*0.5))};
    \addplot {(-0.5-0.6*sin(x)+0.5*sin(2*x*0.5))};
    \addplot {(-0.5-0.7*sin(x)+0.5*sin(2*x*0.5))};
    \legend{$Rk=$\\$0.4$, $0.45$, $0.5$, $0.55$, $0.6$, $0.7$};
    \end{axis}
\end{tikzpicture}
\end{figure}

\begin{figure}[ht] 
\caption{\label{fig:3b} Finding $R_\mathrm{crit}$ in the plot of $\dot{\Theta}_1/\dot{\theta}$ versus $\theta$ for $\dot{\theta} \neq 0$ for different values of $Rk$ when $\alpha = -1/3$.}
\centering
\begin{tikzpicture}
    \begin{axis}[domain=0:179, legend style={cells={align=left},at={(axis cs:160,-0.3)},anchor=south west}, xlabel={$\theta$}, ylabel={$\dot{\Theta}_1/\dot{\theta}$}, no marks,
every axis plot post/.append style={very thick}, grid=both,
    grid style={line width=.1pt, draw=gray!10},
    major grid style={line width=.2pt,draw=gray!50}, y tick label style={
        /pgf/number format/.cd,
        fixed,
        fixed zerofill,
        precision=1,
        /tikz/.cd
    }]
    \addplot {(-(1/3)-0.18*sin(x)+0.5*sin(2*x*(1/3)))};
    \addplot {(-(1/3)-0.2*sin(x)+0.5*sin(2*x*(1/3)))};
    \addplot {(-(1/3)-0.22*sin(x)+0.5*sin(2*x*(1/3)))};
    \addplot {(-(1/3)-0.24*sin(x)+0.5*sin(2*x*(1/3)))};
    \addplot {(-(1/3)-0.26*sin(x)+0.5*sin(2*x*(1/3)))};
    \legend{$Rk=$\\$0.18$, $0.2$, $0.22$, $0.24$, $0.26$};
    \end{axis}
\end{tikzpicture}
\end{figure}
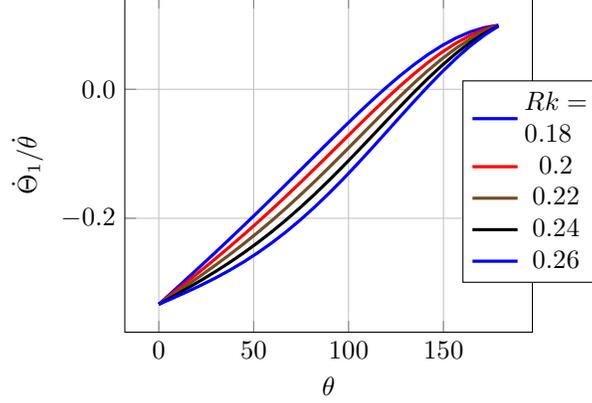

We now carefully examine such values of $R$ that are ``candidates" for $R_\mathrm{crit}$ in the diagram in Fig.~\ref{fig:3}.  We therefore determine that in our model, $R_\mathrm{crit} = 0.5/k$ for $\alpha = -1/2$.  Pretty much in the same manner, we can find for $\alpha=-1/3$ that $R_\mathrm{crit} \approx 0.22/k$ using the diagram in Fig.~\ref{fig:3b}.

\subsection{The critical coupling strength}
The critical coupling strength coefficient $K_\mathrm{critical}$, which in this case, is the threshold for the jump from the state of incoherence to that of bound state AB effect synchronization, can then be calculated as (see, e.g., Ref.~\cite{kuramoto1975self})
\begin{equation}
    K \geq K_\mathrm{critical} := \frac{|\omega_2 - \omega_1|}{2} = |\alpha \dot{\theta}|.
\end{equation}
This further implies that
\begin{equation}
   \frac{|\dot{\theta}|}{2} \geq |\alpha \dot{\theta}|,
\end{equation}
and so the threshold from incoherency to synchronization based on this KM of the bound state AB effect occurs when
\begin{equation}
    |\alpha| \leq \frac{1}{2}. \label{eq:alpha_conditionN}
\end{equation}

This means, for example, if we use an electron as the moving charged particle (the detector), and the source of the magnetic flux is a particle with charge $ne$, then synchronization can occur for $|n|\geq 2$, as specified in Eq.~(\ref{eq:alpha_conditionN}).  The condition in Eq.~(\ref{eq:alpha_conditionN}) confirms that our version of the KM agrees with the physics of the bound state AB effect in that, we have obtained the synchronization condition that coincides with the allowable values of $\alpha$ for the AB effect to occur in the case when the detecting particle is the electron.

\subsection{A more general case}
We now write a more general form of the KM, i.e., when $|\theta_1|$ is not necessarily equal to $|\theta_2|$ ($0 \leq \theta_1 < \pi$ and $-\pi < \theta_2 \leq 0$):
\begin{equation}
 \dot{\Theta}_1 = \left[ \alpha  - R k\sin (\theta_1)  + \frac{1}{2}\sin \left( \Theta_2 - \Theta_1 \right) \right]\dot{\theta}_1,\label{eq:thetadot1withthetaalphageneral}
\end{equation}
and
\begin{equation}
 \dot{\Theta}_2 = \left[ \alpha  - R k\sin (\theta_2)  - \frac{1}{2}\sin \left(\Theta_2 - \Theta_1 \right) \right]\dot{\theta}_2,\label{eq:thetadot2withthetaalphageneral}
\end{equation}
yielding more general conditions for synchronization in terms of the critical coupling strength coefficient $K_\mathrm{critical, \, gen}$:
\begin{eqnarray}
     &&\frac{1}{2}\dot{\theta}_1 =: K_{1} \\
    &&\geq K_{\mathrm{critical, \, gen}} := \frac{1}{2}\Biggl|\alpha \left(\dot{\theta}_2 - \dot{\theta}_1\right) \nonumber\\
    && \quad\quad\quad\quad\quad\quad - R k\dot{\theta}_2\sin (\theta_2) + R k\dot{\theta}_1\sin (\theta_1)\Biggr|,\nonumber
\end{eqnarray}
and
\begin{eqnarray}
     -\frac{1}{2}\dot{\theta}_2 =: K_{2} \geq K_{\mathrm{critical, \, gen}}. 
\end{eqnarray}

In the most general case of $N$ electrons, our version of the KM has the form
\begin{equation}
 \dot{\Theta}_i = \left[ \alpha  - R k\sin (\theta_i)  + \frac{1}{N}\sum_{j=1}^N \sin \left(\Theta_j - \Theta_i \right) \right]\dot{\theta}_i.\label{eq:thetadot1withthetaalphageneralmostgeneral}
\end{equation}

\section{Discussion and Outlook}
We have derived a version of the KM (see Eqs.~(\ref{eq:thetadot1withthetaalpha}) and (\ref{eq:thetadot2withthetaalpha})), which is capable of describing the phase of the bound state AB effect, starting from the expression of the overall wave function for the scattering problem of the quantum system; see, e.g., Section~4 of Ref.~\cite{aharonov1959significance}.

In this model, to observe the AB effect with an electron as the detecting charged particle, for a non-zero angular velocity $\dot{\theta}$ and for $\alpha = -1/2$, the radius $R$ of the circular path through which the electron can orbit around the solenoid, which has a constant magnetic flux $\phi$, should be such that $R < R_\mathrm{crit} = 0.5/k$.  

Notice that, when $\alpha=-1/2$ at $R=R_\mathrm{crit}$, $\dot{\Theta}_i/\dot{\theta}$ becomes constant w.r.t. $\theta$.  We interpret this as the radius through which the magnetic field lines pass, making it magnetically included. When $R>R_\mathrm{crit}$ and $\alpha=-1/2$, the function $\dot{\Theta}_i/\dot{\theta}$ ``flips" w.r.t.~that when $R<R_\mathrm{crit}$, indicating its behavior beyond the magnetic field region.  In this model, the magnetic field region is concentrated at $R_\mathrm{crit}$, which means that the field lines in that area are uniformly vertical, perpendicular to the trajectories of the electrons.  This suggests that for $\alpha=-1/2$, our KM yields the characteristics we desire in the quantum system. For instance, it ensures that the solenoid is sufficiently long and its radius is small enough to create uniform magnetic field lines that pass along its sides at $R_\mathrm{crit}$.  The ``flipping" behavior only confirms that indeed $R_\mathrm{crit}$ is the ``interface'' between the two regions (i.e., less than and beyond the magnetic field) when $\alpha=-1/2$.  In other words, utilizing our KM, we can now understand the magnetically excluded and ``included" regions for $\alpha=-1/2$.  

Although in the case of $\alpha=-1/3$, the graph of the function $\dot{\Theta}_i/\dot{\theta}$ is not as symmetric as that of $\alpha=-1/2$, due to the fact that $\sin(-2\theta\alpha) \neq \sin(\theta)$, however, the radius $R_\mathrm{crit} \approx 0.22/k$ still indicates the magnetic field region; see Figs.~\ref{fig:plot2b} and \ref{fig:3b}.  

One can approximately express the incident wave vector of the electron as $k=2\pi/\lambda_0$, where $\lambda_0$ is the de Broglie wavelength of the incident electron given by (see, e.g., Ref.~\cite{kasunic2019magnetic}) $\lambda_0 = {2\pi\hbar}/({m_0 R|\dot{\theta}_0|}), \quad \dot{\theta}_0 \in \mathbb{R} \setminus \{0\}$, with $m_0$ being the mass of the electron and $\dot{\theta}_0$ its initial angular velocity.  Then for $\alpha=-1/2$, we can have an expression for $R_\mathrm{crit}$ in terms of $\dot{\theta}_0$:
\begin{equation}
    R_\mathrm{crit} = C \sqrt{0.5}  \left( |\dot{\theta}_0| \right)^{-1/2},
\end{equation}
where $C = \sqrt{\hbar /m_0}$.

For any ``allowable" values of $\alpha$ (see, e.g., Table~\ref{tab:1}), we notice that as $Rk \rightarrow 0$, the function $\dot{\Theta}_i/\dot{\theta}$ versus $\theta$ converges into one single plot; see and compare when $Rk=0.0001$ and $Rk=0.02$ in both Figs.~\ref{fig:3} and \ref{fig:3b}. 
 Beyond the magnetic field lines (i.e., $R>R_\mathrm{crit}$), it can be seen that for $\alpha = -1/2$ the function $\dot{\Theta}_i/\dot{\theta}$ at $R=1/k$ is symmetric w.r.t. that as $R \rightarrow 0$; any $R$ larger than this would render behavior that is no longer symmetrical to that as $R \rightarrow 0$.  It is interesting to know, by analyzing further these graphs (or as a further study), how the detecting electron behaves far beyond this region, i.e., $R \gg R_\mathrm{crit}$.
 
On top of this, the condition for the system to undergo a synchronization is $|\alpha| \leq 1/2$, implying that the magnetic flux has the form $\phi = \phi_0 /n$ with $|n| \geq 2$.  Since the synchronization coincides with the condition (see Eq.~(\ref{eq:alpha_conditionN})) when  $\alpha$ yields a detectable AB effect, therefore we say that, in our KM, a synchronization means the detection of the AB effect, while incoherence does not involve AB effect.  We have also made a generalization of our model into $N$ electrons (or in general, $N$ charged particles); see Eq.~(\ref{eq:thetadot1withthetaalphageneralmostgeneral}).

This study also shows that the KM can be derived from a quantum mechanical system with central symmetry (similar to that of the bound state Aharonov-Bohm effect with constant radius and magnetic flux), by expressing its wave function in terms of the incident wave function and the scattered wave function.  

Let $t \in [0,T]$, $T>0$ being the time of interference, $P \in [0,1]$ be the probability of going through different available pathways and $f(P\Theta_2(t) - P\Theta_1(t)) \in \mathbb{R}$ be a function of the ``reduced'' (i.e., $P$-multiple of the) phase difference $(\Theta_2(t) - \Theta_1(t))$ as a function of $t$.  Then, we propose that, from the general expression of the wave function of a quantum mechanical system, which is symmetrical in the polar coordinates, of the form in Eq.~(\ref{eq:psi(t)in_terms_of_inc_and_scattered}), where $\psi_\mathrm{inc} = A(t)\exp\left(-\mathrm{i}\int_0^t \omega_\mathrm{inc}(s)\,\mathrm{d}s\right)$ as before, but that $\psi_\mathrm{scattered} = B_0(t)f(P\Theta_2(T) - P\Theta_1(T))\exp\left(-\mathrm{i}\int_0^t \omega_\mathrm{scattered}(s)\,\mathrm{d}s\right)$, where $B_0(t)f(P\Theta_2(T) - P\Theta_1(T)) = B(t)$ in Eq.~(\ref{eq:psi_scatt_general}), one can formulate a version of the KM of the form
\begin{equation}
    P \dot{\Theta}_i(t) = P \omega_{\mathrm{inc},i} + \frac{P}{N}\omega_{\mathrm{scattered},i}\sum_{j=1}^N f(P\Theta_j(t) - P\Theta_i(t)), \label{eq:PTheta}
\end{equation}
to describe the phase dynamics of the system.  

In our case, we can identify through Eqs.~(\ref{phase}) and (\ref{eq:psi_scattered}) that, $P=1/2$ and $f(P\Theta_2(T) - P\Theta_1(T)) = \sin(\pi\alpha)$.  Therefore, with Eq.~(\ref{eq:PTheta}), we could also have a version of the KM for the bound state AB effect as
\begin{equation}
    \dot{\vartheta}_i = \frac{1}{2} \left[\alpha - R k\sin (\theta_i) + \frac{1}{N}\sum_{j=1}^N \sin(\vartheta_j - \vartheta_i) \right] \dot{\theta}_i, \label{eq:alternative}
\end{equation}
where $\vartheta_i := \Theta_i/2.$  Notice that Eq.~(\ref{eq:alternative}) also produces the same results for the $K_\mathrm{critical}$ and is also equivalent to our main results above, namely, Eq.~(\ref{eq:thetadot1withthetaalphageneralmostgeneral}).

Also in our case, it is important to note that $\omega_{\mathrm{inc},i} = \left(1-{k y_i}/{\alpha} \right) V_{\mathrm{AB},i}$, where $y_i = R\sin (\theta_i)$ and $V_{\mathrm{AB},i} = \alpha \dot{\theta}_i$. Here, $V_{\mathrm{AB},i}$ represents the potential energy of the system; see, e.g., Ref.~\cite{bernido2002path}.

\section*{Acknowledgment}
ARN and RMN-B would like to thank the Department of Science and Technology - Science Education Institute (DOST-SEI) of the Philippine government for financial support in the form of a graduate scholarship. ARN also thanks the Indonesian government for granting the one-year research stay at Universitas Sanata Dharma, Yogyakarta, where this collaborative research work was conducted together with its Mathematics Department of the Faculty of Science and Technology.  JLS was partially supported by the Center for Research in Mathematics and Applications (CIMA) related to the Statistics, Stochastic Processes and Applications (SSPA) group, through the grant UIDB/MAT/04674/2020
of FCT--Funda{\c c\~a}o para a Ci{\^e}ncia e a Tecnologia, Portugal.

\bibliographystyle{plain}

\end{document}